# Neutron tomography of magnetic Majorana fermions in a proximate quantum spin liquid

**Authors:** Arnab Banerjee[1]*, Jiaqiang Yan[2], Johannes Knolle[3], Craig A. Bridges[4], Matthew B. Stone[1], Mark D. Lumsden[1], David G. Mandrus[2], David A. Tennant[5,6], Roderich Moessner[7], Stephen E. Nagler[1]*.

**Affiliations:**

[1]Quantum Condensed Matter Division, Oak Ridge National Laboratory, Oak Ridge, TN 37831, U.S.A.

[2]Material Sciences and Technology Division, Oak Ridge National Laboratory, Oak Ridge, TN, 37831, U.S.A.

[3]Department of Physics, Cavendish Laboratory, JJ Thomson Avenue, Cambridge CB3 0HE, U.K.

[4]Chemical Sciences Division, Oak Ridge National Laboratory, Oak Ridge, TN 37831, U.S.A.

[5]Neutron Sciences Directorate, Oak Ridge National Laboratory, Oak Ridge, TN 37831, U.S.A.

[6]Department of Materials Science and Engineering, University of Tennesee, Knoxville, TN 37996, U.S.A.

[7]Max Planck Institute for the Physics of Complex Systems, D-01187 Dresden, Germany.

**Abstract:**

**Quantum matter provides an effective vacuum out of which arise emergent particles not corresponding to any experimentally detected elementary particle. Topological quantum materials in particular have become a focus of intense research in part because of the remarkable possibility to realize Majorana fermions, with their potential for new, decoherence-free quantum computing architectures. In this paper we undertake a study on high-quality single crystal of α-RuCl$_3$ which has been identified as a material realizing a proximate Kitaev state, a topological quantum state with *magnetic* Majorana fermions. Four-dimensional tomographic reconstruction of dynamical correlations measured using neutrons is uniquely powerful for probing such magnetic states. We discover unusual signals, including an unprecedented column of scattering over a large energy interval around the Brillouin zone center which is remarkably stable with temperature. This is straightforwardly accounted for in terms of the Majorana excitations present in Kitaev's topological quantum spin liquid. Other, more delicate, features in the scattering can be transparently associated with perturbations to an ideal model. This opens a window on emergent magnetic Majorana fermions in correlated materials.**

**Main Text:**

Quantum spin liquids (QSLs) are collective magnetic states that can form in the networks of atomic moments ("spins") in materials. The spins fail to enter an ordinary static ordered state – such as a ferromagnet – as the temperature approaches zero and instead become highly entangled and fluctuate quantum mechanically (1,2). A defining feature of QSLs, connected to their topological nature, is excitations that carry fractional quantum numbers (3) - a phenomenon famously underpinning the physics of the fractional quantum Hall effect (4), magnetic monopoles (5), and spin-charge separation (6). Fractionalization can be seen experimentally by momentum-energy space reconstruction. Inelastic neutron scattering directly probes magnetic correlations in space and time. Our experiments discussed below provide a comprehensive image of the collective magnetic fluctuations in a topological quantum magnet.

Kitaev's QSL (KQSL) (7, 8) is the focus of intense current interest. It occurs in an extremely simple spin network (8, 9) consisting of $S=1/2$ spins on a honeycomb lattice with interaction Hamiltonian:

$$\mathcal{H} = \Sigma_{\gamma,\vec{r}}\left(K^\gamma S^\gamma_{\vec{r}} S^\gamma_{\vec{r}+\vec{\delta}_\gamma}\right) \quad \ldots (1)$$

for either ferromagnetic (FM) or antiferromagnetic (AF) coupling $K$. Here $\vec{r}$ runs over the lattice sites, and the index γ denotes one of the three nearest-neighbor bonds in a honeycomb lattice joined by vector $\vec{\delta}_\gamma$ with Ising interaction strength $K^\gamma$ (See Fig. 1(a)).

The collective excitations of the KQSL comprise gauge fluxes and Majorana fermions (10-12). The latter – originally introduced decades ago in elementary particle physics – are unusual in that they are their own anti-particles; they are ephemeral in that they do not straightforwardly encode a charge or magnetic moment density (unlike familiar particles like electrons or protons). This 'nonlocal' aspect of their existence has led to their proposal as ingredients for robust quantum computation architectures (13, 14). Materials comprising transition metal ions in edge-sharing cubic octahedra with strong spin-orbit coupling, arranged in a layered honeycomb lattice (Fig 1(a)), with weak interlayer interactions (15) are promising candidates for realizing KQSLs. These have included iridates containing $Ir^{4+}$ (16-19), and most recently the $Ru^{3+}$ based honeycomb magnet α-$RuCl_3$ (20-25).

Here we present inelastic neutron scattering on a single crystal of α-$RuCl_3$, providing a complete measurement of the magnetic response function in 4-dimensional energy-momentum space. From a technical perspective, this presents a qualitative advance over polycrystalline samples studied so far (24), or Raman studies (23), which provide a 2-dimensional set of results, as they are unable to distinguish between different scattering directions.

We find a striking broad continuum spectrum of 2-dimensional (2D) magnetic fluctuations centered at the Brillouin zone (BZ) center ($\Gamma$ point), with an energy (E), momentum ($Q$), and temperature (T) dependence lacking a natural explanation by conventional magnetic excitations; however, they strongly resemble the response arising from Majorana fermion excitations in a KQSL. Overall, this spectroscopic tomography of magnetic Majorana fermions allows a direct comparison to theory in significant detail.

A 490 mg single crystal grown via vapor transport of phase-pure $\alpha$-RuCl$_3$ was used. This crystal contains a low incidence of stacking faults and exhibits a single magnetic ordering transition at $T_N$=7 K (see Fig. 1(b), Fig. S1, and 'Materials and Methods' in Supplementary Materials (SM)) (25). Below $T_N$ the individual honeycomb layers exhibit zigzag order, however the ordered moment is small, and, moreover, $T_N$ can be sample dependent (21, 24, 25). This ordering is incidental to the 2D QSL physics of interest on which we concentrate here.

Figure 2 contains a first set of central results. It depicts the temperature and momentum dependence of a magnetic scattering continuum for two energy ranges: 4.5 to 7.5 meV and 7.5 to 12.5 meV. The most salient feature is the robust response centered at the $\Gamma$ point: it is present from low (T=5 K<$T_N$) *all the way* to high (T=120 K >> $T_N$) temperatures of order of the Kitaev coupling estimated at $K^\gamma \approx$70 - 90 K (23, 24). On passing from below to above $T_N$ the central portion of the scattering strengthens. The overall intensity, while weaker, is still readily visible at very high T. At all temperatures this dynamic scattering extends through a large fraction of the Brillouin zone, indicative of short-ranged liquid correlations. (Refer to SM, Fig. S2 and S3 for 2D BZ definitions.)

The energy dependence of the scattering at the $\Gamma$ point is illustrated in Fig. 3(a)/(b) at temperatures below/above $T_N$. Above $T_N$, the broad scattering continuum extends nearly to 15 meV, in keeping with expectations for a pure Kitaev model with $K^\gamma \approx$5.5 - 8 meV (23, 24). Below $T_N$ a fraction of the spectral weight shifts into sharp (i.e., energy resolution limited) spin wave (SW) peaks arising from the small zigzag ordered moments. Crucially, the 2-dimensional nature of the response is shown by the rod-like L dependence of the scattering illustrated in Fig. 3(c)(d).

Most importantly, the persistent energy continuum at the $\Gamma$ point is incompatible with conventional SW physics. Indeed, Fig. 3(e) shows the generic low energy SW response for a zigzag ordered state. This takes the form of dispersive energy-momentum cones centered about each M point magnetic Bragg peak (Fig 3(f)). In SW theory the $\Gamma$ point scattering is present only at certain fixed energy values, unlike the experimentally observed broad energy column (Fig. 2 and Fig. 3(f)). Moreover, SW scattering at long wavelengths is strongly sensitive to cooling through $T_N$ (24), in stark contrast to the continuum. The latter is very broad in energy and almost independent of temperature up to around 100 K (~$K^\gamma$ >> $T_N$) consistent with the thermodynamics of the Kitaev model (22, 27) ($\Gamma$ point scattering at T=120 K is shown in SM Fig. S6(d)). The energy breadth and temperature dependence of the continuum are direct signatures of fractionalized excitations.

Fig. 4(a) shows an extended zone picture of the T = 5 K data integrated between E=[4.5,7.5] meV, symmetrized along the (H,H,0) direction. In addition to the strong scattering at H=K=0, features are now visible near adjacent Γ points ± (1,1), showing that the continuum spectrum repeats every 2nd BZ. Additional scattering at larger Q arises from phonons. In the following we elaborate in more detail the remarkable observation that a Kitaev QSL description reproduces the main qualitative features of the data; in particular, the broad energy width and T-dependence of the scattering continuum, its periodicity and relative orientation in the BZ, which encodes the orientational bond-dependence of the spin anisotropy in Kitaev systems.

The momentum dependence of the scattering for a pure Kitaev model at T=0 is exactly known (11, 12, also see SM, and Fig S5). The dynamical structure factor consists of two energy dependent correlations, those for onsite, $S_0$, and nearest neighbor spins, $S_1$ (See details in SM). For simplicity we compare the scattering to calculations for an isotropic Kitaev model. Although slightly spatially anisotropic Kitaev exchange is likely in α-RuCl$_3$ (21, 25), averaging over the in-plane structural domains (25) reduces its visibility in experiments. Moreover, it is not expected to have a major effect on the higher frequency portion of the collective dynamics discussed here (24). However, in a real material, the effective Hamiltonian includes non-Kitaev terms (9) that extend the liquid correlations and in particular lead to the long range order observed below T$_N$. To date there is no comparably reliable theory available for the response of such an extended Hamiltonian (on which, at any rate, there is not yet a universal agreement for α-RuCl$_3$). As a first, phenomenological, attempt to account for the effect of additional terms, we consider minimally modifying the response function of the pure Kitaev model by varying the ratio of $S_1/S_0$ by a factor '*R*' taken to be momentum independent for simplicity. As shown below, treating this ratio as an adjustable parameter yields an excellent account of the overall momentum dependence of the scattering.

Fig. 4(b) illustrates the scattering for *R*=2 at fixed E=1.2 $K^\gamma$. Remarkably, this calculation captures overall extent, orientation and periodicity of the scattering in reciprocal space. A direct comparison is made in the bottom panel, Fig. 4(c), showing a cut of the intensity as a function of momentum along the (H,H) direction, integrated over a narrow band around |(K,-K)|=0. Also shown are three model calculations for an isotropic Kitaev model at fixed E=1.2 $K^\gamma$: AF (violet), FM (green), and AF response modified using *R*=2 (red). The FM model, in the absence of any further terms in the Hamiltonian, is clearly incompatible with the data as it shows a local minimum at the Γ point. The unmodified AF Kitaev response gives a reasonable description of the data but fails to capture the full intensity variation (see details, and Fig S5 in SM). The modified AF Kitaev response fits the data best with $R \approx 2$ indicating a relative enhancement of the spatial correlations.

The results reported here provide a unique picture of the magnetic response function of α–RuCl$_3$ in momentum-energy space, and demonstrate unequivocally the presence of a strong quantum fluctuation continuum centered at the Γ point. The continuum response is grossly incompatible with SW theory, and indeed defies any known explanation in terms of conventional dispersive spin flip or single particle magnetic excitations. Instead, the central features of the continuum are well described by the scattering for an AF KQSL; with one phenomenological fitting parameter, nearly quantitative agreement is obtained. This even leads to an intuitive physical interpretation in terms of the fractionalized degrees of freedom because the exact response function of the pure Kitaev model directly reflects the Majorana fermion density of states in the presence of a static heavy pair of emergent fluxes (11, 12).

One feature of the data that is not well-described by a pure Kitaev model is the six-pointed star shape of the scattering in reciprocal space. However it can be shown (see SM) that modest amounts of additional neighbor correlation or simple perturbations based on mean-field approaches (28) away from the integrable model can yield a similar shape even in the disordered state.

The data presented here is a significant step in developing a complete understanding of the low and high energy dynamics in α-RuCl$_3$. The good agreement of the continuum scattering with the simple AF KQSL is complementary to current DFT calculations relating the low energy spin ½ description of the material to details of the electronic structure (29, 30). Further effort is needed to converge on an explanation of the sign of the Kitaev interaction, and to determine the magnitude of additional interactions. It would be of great interest to develop a theory that describes concomitantly both the low-energy response of the ordered state and the broad quantum fluctuation continuum. At the same time, the evident proximity of the system to a true KQSL is a strong incentive for exploring the effects of doping, pressure and field to determine a full picture of the ground and excited states. With this work, a comprehensive measurement of the high-energy excitations is now available to the community in a proximate Kitaev material of choice, and opens up the opportunity to investigate the magnetic version of the intriguing and elusive Majorana fermions.

———————

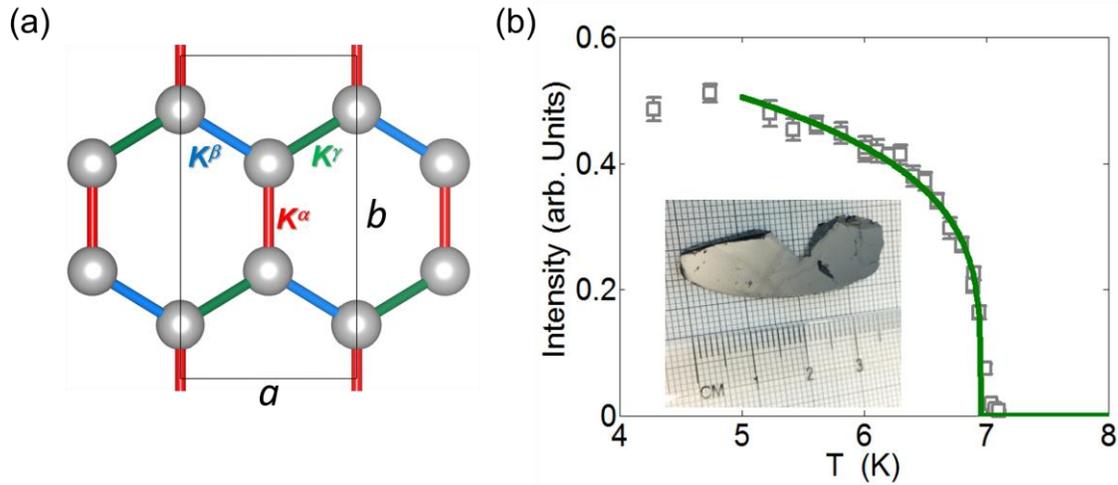

**Fig. 1: Structure and magnetism in single-crystal α-RuCl₃** (a) The honeycomb lattice of $Ru^{3+}$ magnetic ions in one plane of α-RuCl₃ showing the projections of the three mutually competing Ising bonds corresponding to the Kitaev terms in equation (1). (b) The intensity of the magnetic Bragg peak occurring at the M point of the 2D honeycomb lattice corresponding to a zigzag structure with three-layer stacking. The single sharp magnetic transition is characteristic of crystals with few or no stacking faults (25). The solid line is a power-law fit yielding $T_N = 6.96 \pm 0.02$ K and a critical exponent $\beta = 0.125 \pm 0.015$, suggesting 2D Ising behavior. (Inset) The 490 mg single-crystal of α-RuCl₃ used for the neutron measurements.

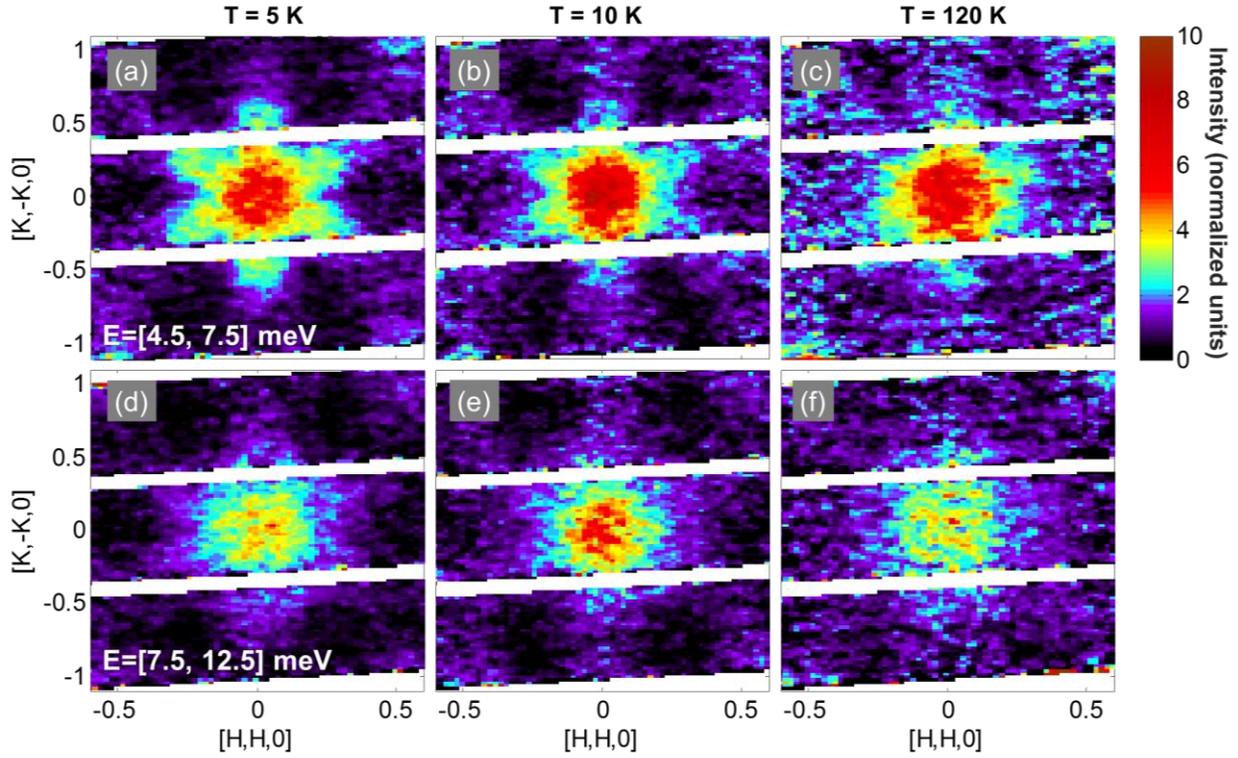

**Fig. 2: Momentum and temperature dependence of the scattering continuum:** Neutron scattering measurements using fixed incident energy $E_i$ = 40 meV, projected on the reciprocal honeycomb plane defined by the perpendicular directions (H,H,0) and (K,-K,0), integrated over the interval L=[-2.5, 2.5]. Intensities are denoted by color as indicated in the scale at right. Measurements integrated over the energy range [4.5, 7.5] meV are shown on the top row at temperatures (a) 5 K, (b) 10 K, and (c) 120 K. The corresponding measurements integrated over the interval [7.5, 12.5] meV are shown in panels (d), (e), and (f). (The white regions lack detector coverage.)

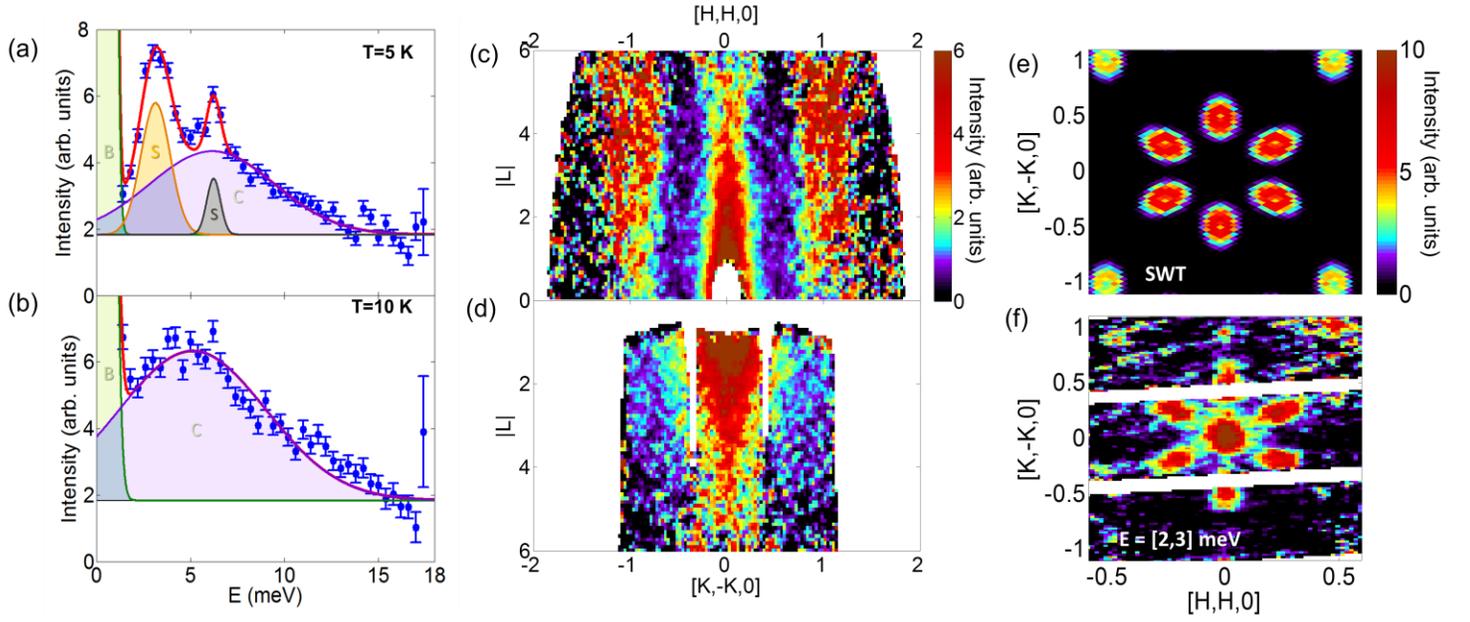

**Fig. 3: Detailed features of the Γ point scattering (see text):** (a),(b) The energy dependence of the scattering integrated over the constant momentum volume defined by the following integration ranges: L = [-2.5,2.5]: [ζ,0] ≡ (K,-K,0) over the range ζ = [-√3/10,√3/10]: [ξ,0] ≡ (H,H,0) over the range ξ = [-0.1,0.1], at temperatures (a) 5 K, and (b) 10 K. The solid lines are guides to the eye composed of fits to Gaussian peaks: "E" represents an elastic contribution, "S" spin wave peaks appearing below $T_N$, and "C" the continuum. (c) Scattering symmetrized in the (H,H,L) plane and over positive and negative L, integrated over the intervals ζ =[-√3/10,√3/10], and E=[4.5,7.5] meV. (d) Scattering in the (K,-K, L) plane integrated over ξ = [-0.1,0.1] and E=[4.5,7.5] meV. (e) Representative low-energy spin wave scattering expected for a zigzag ordered phase. (f) Scattering at a temperature 5 K integrated over L=[-2.5,2.5] and E=[2,3] meV. (The white regions lack detector coverage.)

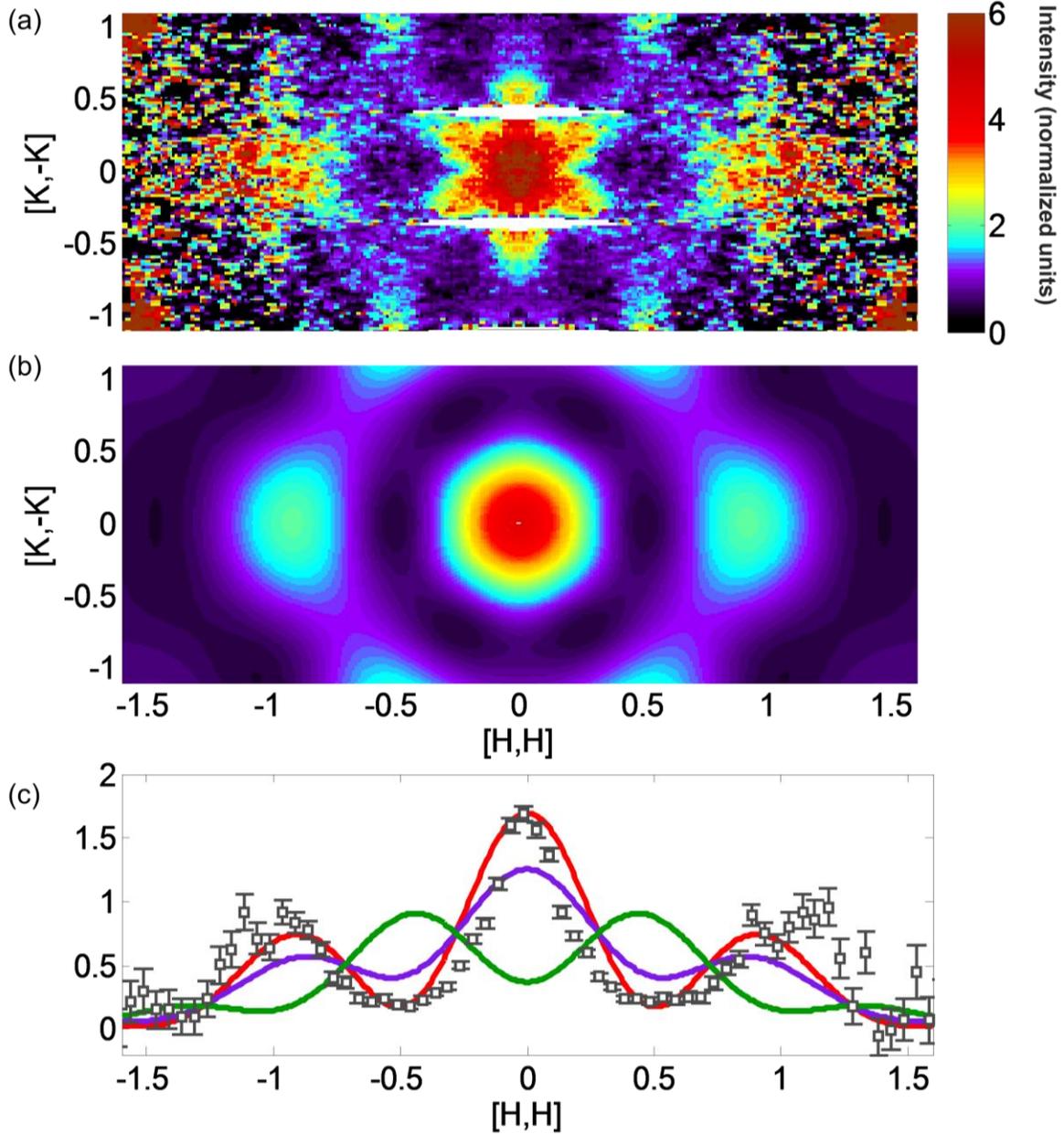

**Fig. 4: Comparison of the scattering with Kitaev model calculations:** (a) The data at $E_i$=40 meV, T=10 K integrated over range E= [4.5,7.5] meV and L = [-2.5,2.5] and symmetrized along the (H,H) direction. (b) The expected scattering from an isotropic AF Kitaev model at an energy $E = 1.2\ K^\gamma$, taking into account the neutron polarization and the $Ru^{3+}$ form factors. (c) Plot of the non-symmetrized data (points with error bars) along (H,H,0) at T =10 K, integrated over the same L and E intervals as (a) as well as $\zeta$ = [-√3/10,√3/10]. The solid red line is the calculated scattering for an AF Kitaev model with $R = 2$ as discussed in the text. The solid violet line represents the corresponding unmodified AF Kitaev model, and the green line the FM Kitaev model. Some of the scattering at larger Q near (H,H) = ±(1,1) is due to phonons.

**Acknowledgements:**

The authors acknowledge valuable discussions with C. Batista, K. Burch, H. Cao, B. Chakoumakos, G. Jackeli, G. Khalliulin, J. Leiner, R. Valenti and S. Winter. J. K. and R. M. particularly thank J. Chalker and D. Kovrizhin for collaboration on closely related work. We thank O. Garlea for assistance with the measurement on HYSPEC. The work at ORNL's Spallation Neutron Source was supported by the United States Department of Energy (US-DOE), Office of Science, Basic Energy Sciences (BES), Scientific User Facilities Division. Part of the research was supported by the US-DOE, Office of Science, Basic Energy Sciences, Materials Sciences and Engineering Division, under contract number DE-AC05-00OR22725 (JY, CB and DM). DM was funded in part by the Gordon and Betty Moore Foundation's EPiQS Initiative through Grant GBMF4416. The work at Dresden was in part supported by DFG grant SFB 1143 (JK and RM), and by a fellowship within the Postdoc-Program of the German Academic Exchange Service (DAAD) (JK). The collaboration as a whole was supported by the Helmholtz Virtual Institute "New States of Matter and their Excitations" initiative.

# Supplementary Materials:

## A. Materials and Methods:

**(i) Sample preparation and characterization:** Commercial RuCl$_3$ powder procured from Sigma-Aldrich was purified in-house. The resulting powder is better than 99.9 % pure RuCl$_3$, confirmed using Inductive-Coupled Plasma mass spectroscopy (Galbraith Laboratories, Inc.) and X-Ray diffraction using PANalytical Empyrean X-ray diffractometer, consistent with the earlier results (24). Single crystals of α-RuCl$_3$ were grown from this purified powder using the vapor-transport technique. All the single-crystal data shown in this manuscript as well as reference (25) was collected on samples from the same growth batch. The temperature dependence of the magnetic susceptibility (Fig. S1) was obtained using a Quantum Design magnetic property measurement system (MPMS) in the temperature range of 2K-320K. This shows just one magnetic transition at $T_N$ = 7 K (see Fig S1 inset) consistent with reference (25). The main panel of Fig. S1 shows the inverse susceptibility (1/χ) measured with field in and out-of-plane. A linear fit in the high-temperature limit (T = 170 – 320 K) shows that these have similar slopes corresponding to an effective high-temperature moment of roughly μ=2.35 μ$_B$. Below 120 K (shown by black arrow) the in-plane inverse susceptibility deviates from linearity, however the out-of-plane susceptibility does not show this behavior. The extrapolated Curie-Weiss temperatures (Fig.S1 caption) are overall close to recently published literature on bulk measurement in single-crystal α-RuCl$_3$ (31, 32).

**(ii) Neutron diffraction and 4D inelastic neutron tomography:** All neutron scattering was performed on one single piece of crystal about 490 mg in mass and 1.5 × 4.0 × 0.1 cm$^3$ in size. This was sealed in a thin walled aluminum canister with 1 atmosphere Helium gas in order to both avoid moisture and provide a thermal anchor. The low T crystal structure of the sample complies with the *C*2/*m* space-group symmetry below consistent with previous reports (21, 25, 33). Neutron diffraction measured at the Spallation Neutron Source (SNS) HYSPEC (34) instrument ($E_i$ = 15 meV, Fermi chopper spinning at 360 Hz) showed that the low T magnetic order was consistent with reference (25): an in-plane zigzag structure with a 3 layer stacking periodicity. The sample showed a single magnetic transition at 7 K and no evidence for ABAB type stacking contamination.

Inelastic neutron scattering measurements were performed using the SEQUOIA (34, 35) chopper spectrometer at the SNS. The sample can was mounted in a closed-cycle Helium based refrigerator for temperature control. An incident energy of $E_i$ = 40 meV (λ = 1.43 Å) was used with the $T_0$ chopper at 60 Hz and Fermi chopper at 360 Hz (34). The resolution of the instrument with this setting is 1.10 meV FWHM (full width at half maximum) at the elastic position and 0.96 FWHM meV at 6 meV energy transfer. The 2D detectors at SEQUOIA covered an angular range of up to 54° in the horizontal plane and ±18° in the vertical plane corresponding to an overall solid angle of 0.863 steradians (35).

For ease of discussion and consistency with the 2D honeycomb lattice the Q dependence of the inelastic scattering in this paper is plotted using the notation of the trigonal space-group (*P*3$_1$12, *a=b*=5.975 Å, *c*=17 Å). In this notation, the reduced data is plotted along the orthogonal axes (1,1,0), (1,-1,0) and (0,0,1), denoted in this paper by (H,H,0), (K,-K,0) and (0,0,L). The (H,H,0) axis is in units of 2.10 Å$^{-1}$, the (K,-K,0) axis is in units of 1.21 Å$^{-1}$, and the (0,0,L) axis is in units of 0.37 Å$^{-1}$. Note that the corresponding coordinates in the orthogonal *C*2/*m* space group symmetry (21, 25) would be (2,0,0),(0,2,0) and (0,0,1/3) respectively.

For the inelastic measurements the crystal was mounted with the (H,H,0) and (0,0,L) axes in the horizontal plane, and the orthogonal (K,-K,0) axis pointing vertically upwards. Tomographic reconstruction in the 3 momentum directions was performed by rotating the crystal about the (K,-K,0) axis over 360° (in 2.5° steps for 5 K data, and 5° steps for 10 K and 120 K data), measuring for a fixed amount of proton charge on the spallation target (rougly15 minutes per measurement). This provides a continuous coverage over all the three orthogonal momentum-transfer dimensions (H,H,0), (K,-K,0) and (0,0,L). The 4$^{th}$ dimension of energy-transfer was obtained via the time-of-flight of the neutrons. The individual measurements at each rotation angle were normalized for the proton charge on the spallation target, corrected for detector sensitivity using Vanadium normalization, and then binned from 4D laboratory coordinates to 4D sample coordinates using standard direct geometry chopper spectrometer reduction routines embedded within the Mantid software (36) with an energy bin of 0.4 meV and momentum bin of 0.04 Å$^{-1}$  Data from both the sample and the empty canister background were reduced in an identical fashion as above.

**(iii) Data analysis:** The data was subtracted of the background, rebinned, and projections along the appropriate crystallographic axes were made for presentation and analysis purposes using MSlice distributed by the DAVE project (37) and HORACE software distributed by ISIS (38). The plots were made using Matlab. The spin wave calculation in Fig. 3(e) and Fig. S6(a-c)  was performed in Matlab using SpinW package (39). The neutron intensities derived for the Kitaev model started from the exact dynamical structure factor calculated as described in references (11) and (12) and accounted for neutron polarization terms and the magnetic form factor of  Ru$^{3+}$ in the same manner as described in reference (24), with some additional details presented below in the Supplementary Materials section on 'Neutron scattering cross section on pure Kitaev model'.

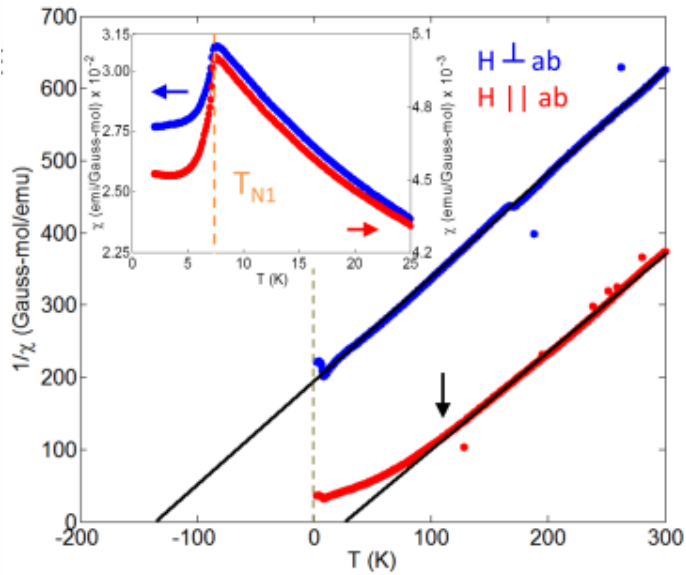

**Figure S1:** Inverse susceptibility of pure α-RuCl$_3$ measured with H∥ab (red) and H∥c (blue) is fitted to a linear behavior in the regime 150 K – 300 K to yield an effective moment size of µ=2.39(2) µ$_B$ and µ=2.33(2) µ$_B$ for in-plane and out-of-plane measurements respectively with the extrapolated Curie-Weiss temperatures are θ$_c$ = -130(4) K for χ∥c and θ$_{ab}$ = 32(3) K for χ∥ab.

## B. Real and reciprocal space definitions:

Although at low temperatures the crystals show a monoclinic *C2/m* space group (21, 25), the honeycomb lattice of $Ru^{3+}$ atoms in each plane is symmetric to <0.2 % (25). Moreover, there are three possible monoclinic domains related by 120° rotations. The in-plane dependence of the inelastic scattering is best depicted using the reciprocal space appropriate for the honeycomb lattice, which is the same as the in-plane reciprocal lattice of the trigonal space group structure.

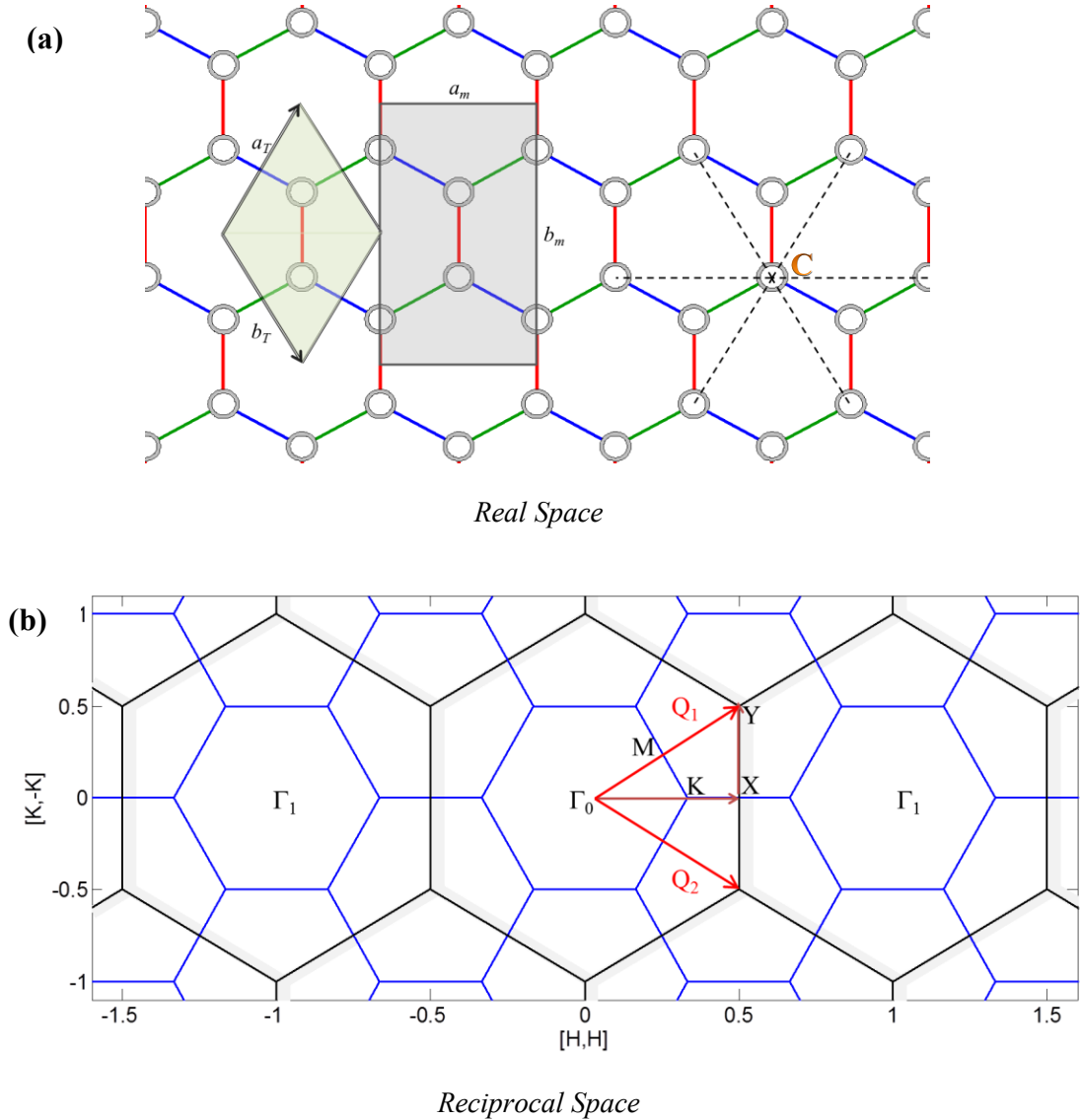

*Real Space*

*Reciprocal Space*

**Figure S2:** (a) The honeycomb lattice showing the locations of the $Ru^{3+}$ (empty circles). The unit cells for the trigonal and the orthorhombic space groups are shaded. $a_T$ and $b_T$ are the lattice vectors in a trigonal space-group as used in this paper, while $a_m$ and $b_m$ are the same for the monoclinic space group (21, 25, 33). The black dashed lines show the 6 next nearest neighbors for the atom in the center marked 'C'. (b) The reciprocal lattice of the honeycomb lattice in the trigonal space group.

We nevertheless note that all calculations for the pure Kitaev model and the data can be similarly reproduced using a *C2/m* type space group, and the conclusions presented in this paper hold regardless. In Fig S2(a), we show the in-plane real space lattice vectors for trigonal and monoclinic structures; the honeycomb reciprocal space is shown in S2B. For ease of comparison, the reciprocal space diagram is plotted over the same range as in Fig. 4(a)(b).

Figure S3 shows the elastic scattering data (E = [-0.25, 0.25] meV) taken with $E_i$ = 25 meV, at T=5 K < $T_N$. This shows the magnetic Bragg peaks in the HK0 plane, plotted over the same range as in fig. 2 of main text. The reciprocal space BZ diagram is superimposed to give a perspective on the extent of the data. The reciprocal lattice vectors are marked in black arrows. Magnetic Bragg peaks (faint green or red dots) appear at the M-points ((1/2,0,L) type points), while the structural Bragg peaks (intense red dots) appear at the (1,0,L) type points.

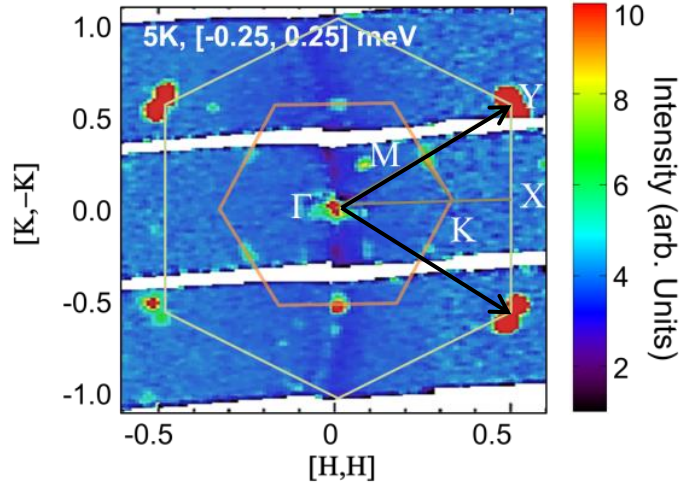

**Figure S3:** The same reciprocal space image of Fig. S2 when superimposed on top of the data taken at SEQUOIA with $E_i$ = 25 meV, T = 5 K, integrated in E = [-0.25,0.25] and L = [-2,2]. The smaller orange hexagon and the larger grey hexagon represent the first and the second Brillouin Zones respectively. The magnetic Bragg peaks are apparent as green dots at the M-points.

**C. The neutron scattering cross section calculations for the pure Kitaev model:**

The neutron scattering cross section calculations for the pure Kitaev model needed to compare to the experimental data are obtained by utilizing the known exact calculations of the zero-temperature dynamic response functions (11, 12), and subsequently accounting for the neutron polarization factors and the $Ru^{3+}$ form factor (24).

We outline some aspects of the exact calculation for the isotropic Kitaev model below, for details see (11, 12). In the Kitaev model spins fractionalize into static $Z_2$ gauge fluxes and itinerant Majorana fermions. The dynamic magnetic response function can be expressed as a Fourier transform of the spin-spin correlation function (following equation 2 in 12, adopting a Pauli Matrix notation):

$$S^{aa}(\vec{q},\omega) = \frac{1}{N}\Sigma_{ij}\exp(-i\vec{q}\cdot\vec{r}_{ij}^{a})\int_{-\infty}^{\infty}\langle\hat{\sigma}_i^a(t)\hat{\sigma}_j^a(t)\rangle\exp(i\omega t)\,dt \qquad \ldots \text{(SE1)}$$

Here $\vec{r}^a_{i,j}$ is the vector connecting spin-1/2 degrees of freedom between nearest neighbor (NN) sites $i$ and $j$ for the bond type $a=\alpha,\beta$ or $\gamma$ of the honeycomb lattice as explained in Fig. 1(a) of the main text. The calculation of the correlation function is facilitated by the fact that fluxes are non-dynamical -- the action of a spin operator inserts a static nearest-neighbor (NN) flux pair. For a nonzero matrix element the second spin operator acting at a later time needs to remove the same fluxes, leading to ultra-short ranged correlations in real space, e.g. $S_{ij}=0$ for $<i,j>$ any further than NN apart (10). For the remaining onsite correlators $S_0 = S_{ii}^{aa}(\omega)$ and NN correlators $S_1 = S_{ij}^{aa}(\omega)$ (independent of $aa$ for isotropic couplings) it is possible to obtain an expression in terms of Majorana fermions (11). This takes the form of a local quantum quench of Majoranas which can be solved numerically exactly even in the thermodynamic limit (11, 12). The resulting functions $S_0$ and $S_1$ are continuous functions of energy. While the onsite correlation $S_0$ is always positive as a function of frequency the NN component $S_1$ changes sign above $\omega \sim 0.8$ K as shown in figure S4.

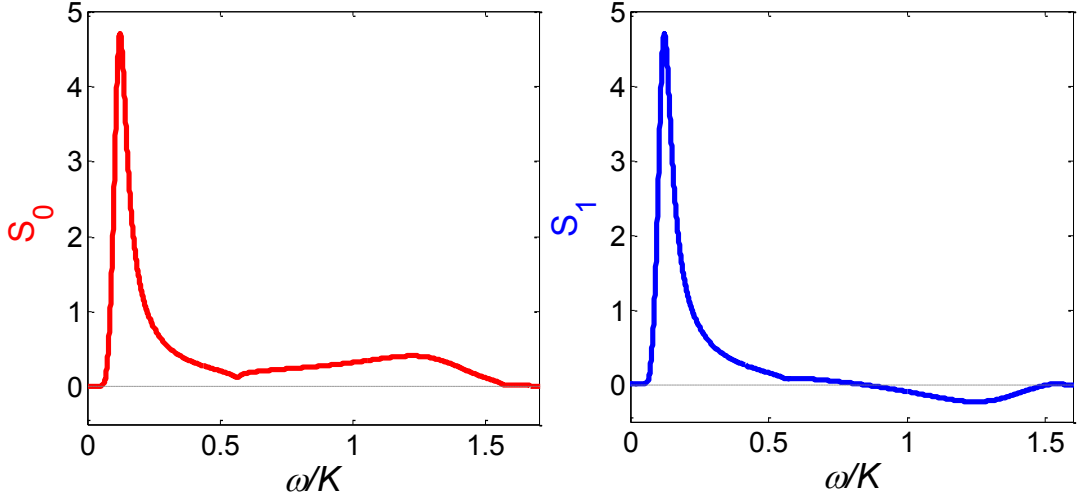

**Figure S4:** The functions $S_0$ and $S_1$ plotted versus $\omega$ (in units of the Kitaev energy $K$).

Substituting these results into equation SE1 yields:

$$S^{aa}(\vec{q},\omega) = 2(S_0(\omega) - sgn(K)\cos(\vec{q}\cdot\vec{R}_{NN}^a)\cdot S_1(\omega)) \qquad \ldots \text{(SE2)}$$

Here the nearest neighbor correlator is given by each type of honeycomb bond $\vec{R}_{NN}^a = \vec{r}_i - \vec{r}_j$. The term $sgn(K) = +1$ for AFM Kitaev interactions and $-1$ for FM Kitaev interactions (11, 12). Eq. SE2 implies that over the broad region near $E \approx K$ where $S_1$ is negative the AF response has a peak at the $\Gamma$ point, while the FM response has a local minimum. (Note: We are setting $\hbar = 1$ so that E and $\omega$ are used interchangeably here).

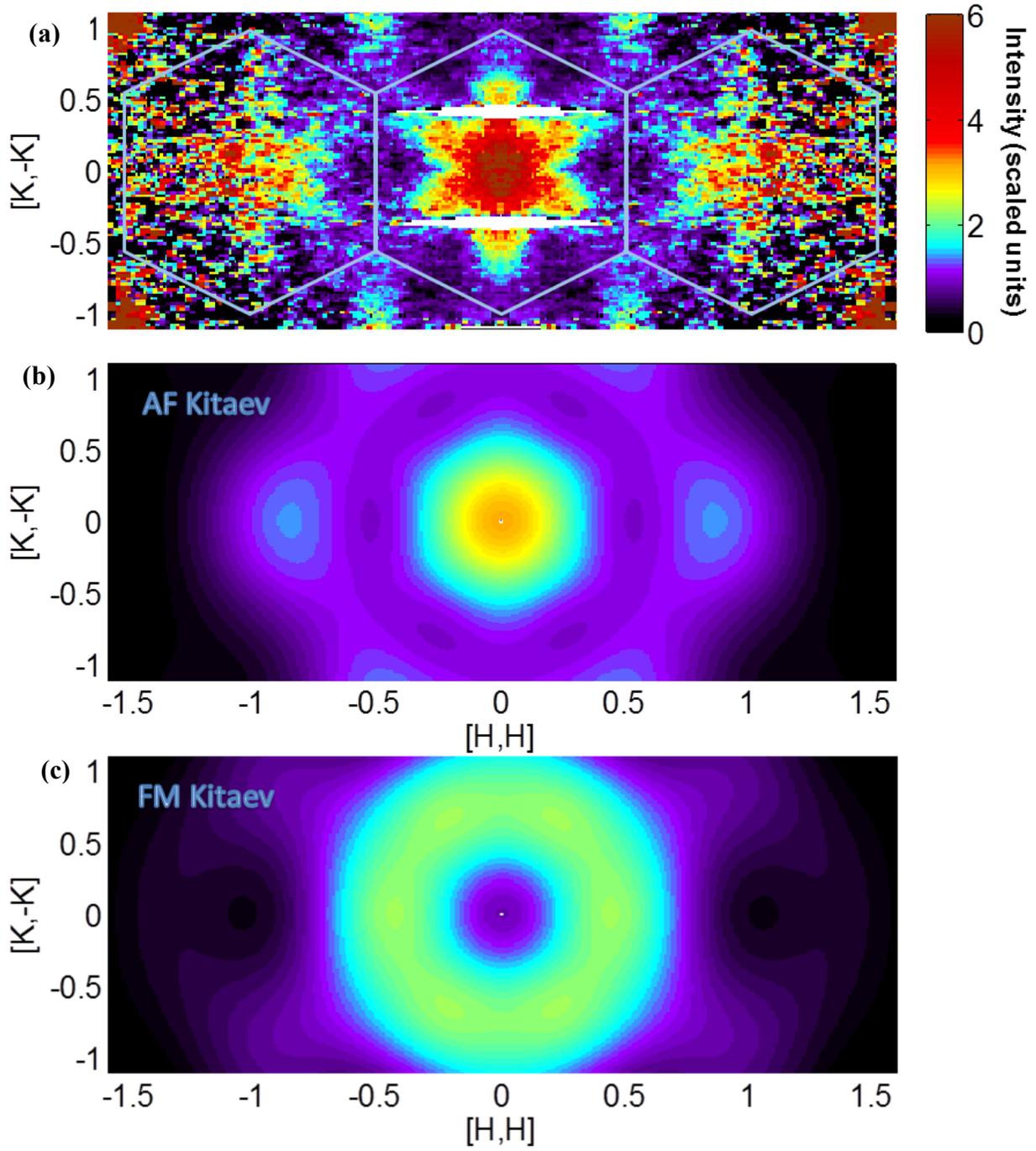

**Figure S5: (a)** The data from Fig 4a of main text. The 2$^{nd}$ BZ is superimposed (light blue lines) **(b,c):** The intensity distribution of the pure Kitaev calculation with E = 1.2 $K$ for the AF Kitaev model **(b)** and the FM Kitaev model **(c)** including the neutron polarization factor and the Ru$^{3+}$ form factors.

In order make contact with the experimental neutron scattering cross-section one needs to take into account the neutron polarization factors and the form factor (40). This leads to an expression for the scattering intensity:

$$I(\vec{q},\omega) \propto |F(q)|^2 \sum_{a=\alpha,\beta,\gamma} \left\{ S^{aa}(\vec{q},\omega)\left(1-\left(\frac{q^a}{q}\right)^2\right) \right\} \quad \ldots(SE4)$$

where $q^a$ is the projection of the momentum vector on the direction of the spin components, and $F(q)$ is the (assumed isotropic) magnetic form factor. The mechanics of the calculation utilizes spin axes where the <111> direction in spin space is perpendicular to the honeycomb plane, and the projections of the <100>, <010>, and <001> directions are naturally separated by 120° in the plane (15).

The result of SE4 plotted in the *ab*-plane is shown for AFM Kitaev interactions in Fig. S5(b) and for FM Kitaev interactions in Fig. S5(c). For a visual reference, Fig 4(a) from the main text is also presented at the top.

The exact solution of the Kitaev model captures the qualitative features of the data, e.g. the symmetry and periodicity in the BZ and, most importantly, a maximum of intensity around the Γ point extending over a large fraction of the BZ and a very broad energy range. For the AF Kitaev model, as a function of energy, a local intensity maximum close to the Γ point occurs near $E = 1.2\ K$, which is similar to the location of the maximum in the powder averaged model (24). When comparing to experiment there are quantitative differences from the Kitaev model expected to arise from the additional interactions which are also responsible for the low temperature magnetic long range order. As a first step to account for this empirically, keeping the assumption of short-ranged NN spin correlations only, we introduce a single multiplicative fitting parameter, $C_1$ on the NN correlator $S_1$,

$$I(\vec{q},\omega,C_1) \propto |F(q)|^2 \sum_{a=\alpha,\beta,\gamma} \left\{ \left(S_0(\omega) - C_1 sgn(K)\cos(\vec{q}\cdot\vec{R}_{NN}^a)\cdot S_1(\omega)\right)\left(1-\left(\frac{q^a}{q}\right)^2\right) \right\} \ldots (SE5)$$

This equation is fit to the data of the cut shown in fig. 4(c) in the main text, for $E = 1.2\ K$ we obtain $C_1 = 2.0 \pm 0.1$. In fact the fitted value of $C_1$ is not very sensitive to the value of $E/K$ over the range $1 \leq E/K \leq 1.4$. This indicates that the ratio of the onsite to NN correlator in α-RuCl$_3$ is enhanced by a factor of roughly 2 (Note, this is same as the factor '$R$' in the main text).

We note that one expects the low frequency dynamics of the material to be strongly affected by smaller non-Kitaev terms in the Hamiltonian, and it is not appropriate to compare data in that regime to the calculation above*. In the ordered states one observes spin waves that are seen to disappear at temperatures just above $T_N$. Nevertheless it is interesting to consider whether spin waves can explain a continuum response at the Γ point. We have calculated spin waves for many different model Hamiltonians with $K, J_1, J_2, Γ$, (some of which are described in Supplementary Information of (24)). Although many allow for some response at the fixed energies at the Γ point, we have not found any that yield the correct scattering over a broad energy continuum. As an example, the DFT (30, 31) and some of the quantum chemistry calculations (42, 43) performed by various groups using inputs from structure and susceptibility measurements (21, 25, 32) and references therein, predict a FM Kitaev term $K \sim -Γ$. A plot of the classical spin wave dispersion relation for a representative set of parameters ($K=-Γ=-6$ meV, $J_1=-$ 0.75 meV, $J_3 = 0.0^+$, summed over the three 120° domains) in that regime is presented in Fig. S6(a-c). Although at lower energies it seems to be roughly consistent with the data at the M points, the in-plane intensity distribution for higher energies (for e.g. E > 3 meV) has no intensity at the Γ-point. At any rate, our data shows that the scattering at the Γ-point extends to 10-15 meV and is strongly present even at T = 120 K, much higher than $T_N$ (and hence beyond the purview of any spin wave calculation), as shown in Fig. S6(d).

*A recent study of the role of small perturbations from the Kitaev point (41) shows how the response function is modified particularly at low energies, i.e. below the flux gap. The relevant energy window is not accessible in the current experiment.

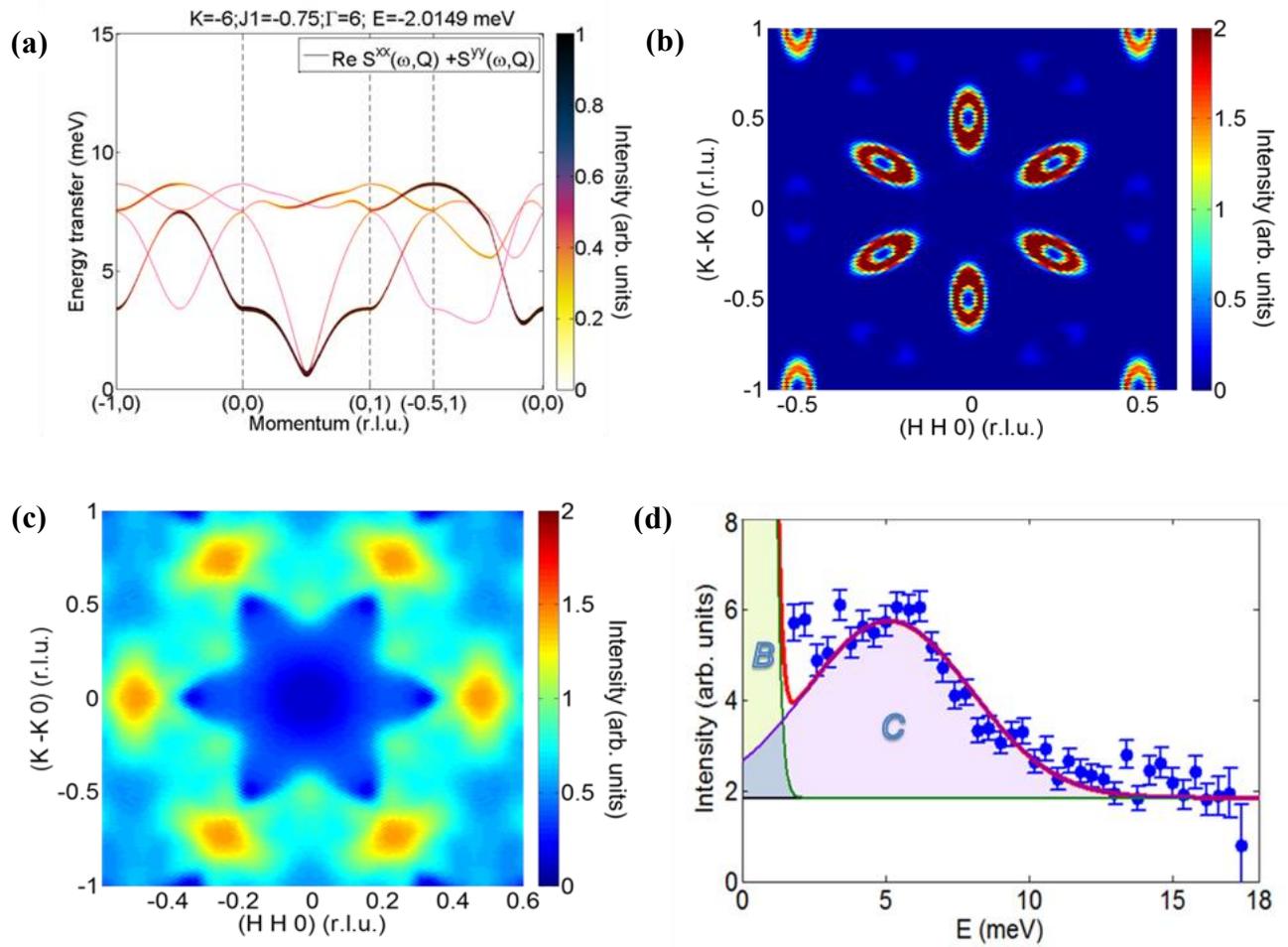

**Figure S6: (a)** The SWT dispersions of a FM Kitaev model with parameters $K=-\Gamma=-6$ meV and $J_1=-0.75$ meV. **(b)** Calculated SWT intensity slice at E = [1.5, 2.0]. **(c)** The constant energy slice at a higher energy E = [6, 10] meV lacks significant intensity at the Γ-point. **(d)** The cut through the $E_i$ = 40 meV, T=120 K data shows that the scattering is persistent up to temperature several times of $T_N$ at the Γ-point. ('B' is the scattering from the elastic line and 'C' is the continuum as defined for Fig 3(a)(b)).

**D. Extending the pure Kitaev solution by adding an adhoc next-nearest neighbor (NNN) interaction:**

While many aspects of the high-energy mode are captured by the pure Kitaev model, it does not reproduce the detailed in-plane Q dependence, specifically the six-pointed star-shaped intensity at the Γ-point (Fig 4(a)). Beyond the possibility that NN correlations are enhanced by non-Kitaev terms, generically non-vanishing longer-range correlators will also be induced. Therefore, here we consider the addition of an adhoc NNN correlator added as a simple fraction $C_2$ of the NN correlator:

$$I_{NNN}(\vec{q}, \omega, C_1, C_2) \propto C_2 \sum_b \left(\cos(\vec{q} \cdot d^b_{NNN}) \cdot C_1 S_1(\omega)\right)\left(1 - \left(\frac{\vec{q} \cdot d^b_{NNN}}{\vec{q}}\right)^2\right) \quad \ldots \text{(SE6)}$$

Here $b$ sums over the 6 NNN terms for a given $Ru^{3+}$ atom shown in fig. S2(b). The final intensity is then SE6 added to:

$$I(q, \omega, C_1, C_2) \propto I(\vec{q}, \omega, C_1) + I_{NNN}(\vec{q}, \omega, C_1, C_2) \quad \ldots \text{(SE7)}$$

In Fig. S7 we show the result for $C_2 = -0.1$ (maintaining the $S_1$ multiplier $C_1 = 2$), where the NNN correlator is 10 % of the enhanced NN correlator but with an opposite sign, which successfully reproduces the star-like shape in the first BZ. This simplistic treatment fails to satisfactorily capture the scattering beyond just the first BZ. It nevertheless shows that correlations that are somewhat extended in real space (compared to the ultra-short ranged Kitaev QSL) account naturally and simply for the star shape.

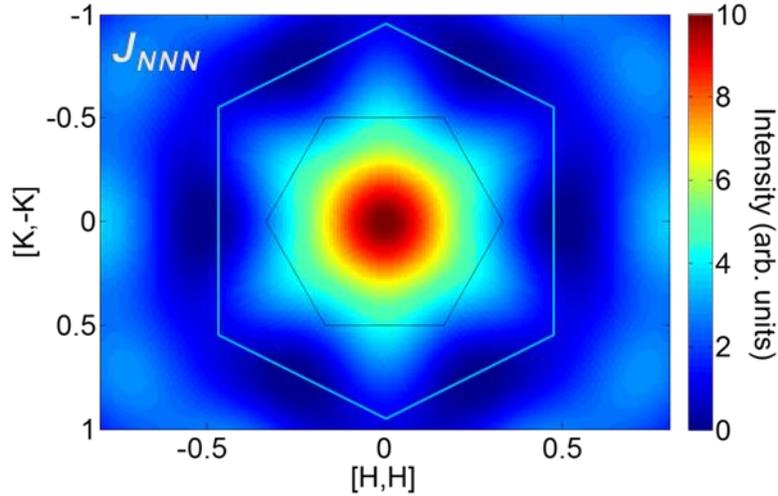

**Figure S7:** The pure Kitaev calculation with a next-nearest neighbor correlator as described in text, resemble a star pattern at the Γ-point.

### E. A mean field extension of the pure Kitaev solution:

In order to account for the tendency of the material to order at low temperatures, we use a phenomenological mean-field model that allows for a calculation of the response function within the framework of the Random Phase Approximation (RPA). Following earlier work on coupled chains (28), and knowing empirically that α-RuCl₃ forms a zigzag ordered state at low temperatures, we introduce an effective symmetry-breaking zigzag field $J_h$ within a RPA. This modifies the magnetic susceptibility such that

$$\chi_{RPA}(q,\omega) = \frac{\chi(q,\omega)}{1-2c_h J_h(q)\chi(q,\omega)} \quad \ldots \text{(SE8)}$$

where $c_h$ is a prefactor capturing the strength of $J_h$.

The RPA calculations start from the complex susceptibility χ, where $Im(\chi)$ is directly proportional to the spin structure factor $S^{aa}(\vec{q},\omega)$ and $Re(\chi)$ is obtained using the Kramer-Kronig relations. To incorporate the short-range correlations of the zigzag order into $J_h$ we work with the simplest motif of the honeycomb lattice consisting of just three NN bonds as shown in Fig. S8 – the "Y" motif containing four spins. For the zigzag ground state three of these spins will have the same sign and one at the corner will have an opposite sign. The zigzag field at the central atom of the motif is then given by:

$$J_h(q) = \sum_{k=4 \text{ atoms}}(\sigma_k e^{iqd_k}) \quad \ldots \text{(SE8)}$$

where $\sigma_\kappa$ takes a value of +1 for an up-spin and -1 for down-spin.

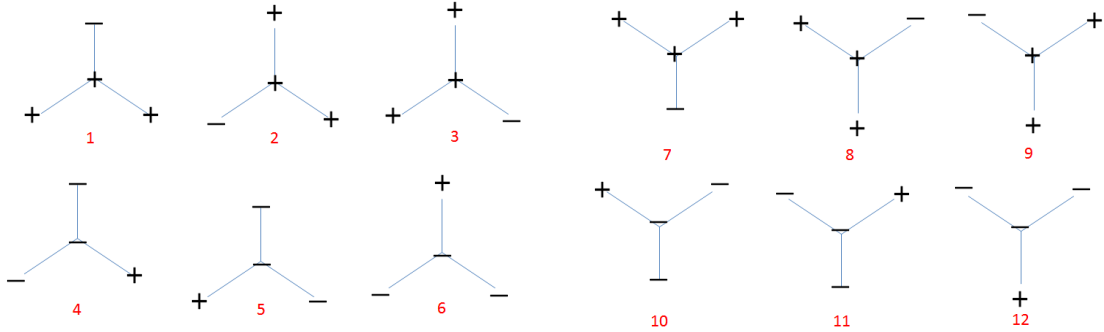

**Figure S8:** The 12 possibilities summed up for RPA to preserve symmetry. Details in text.

We do not want to explicitly break translational or spin-rotational invariance because we are only concerned with data above $T_N$. Therefore we symmetrize all possible zigzag configurations leading to the averaging of the 12 diagrams shown in Fig. S8. The final result is shown in Fig. S9. For $c_h > 0.2$ the original AF Kitaev intensity (i.e., maintaining $C_l=1$) starts to get visibly modified, with a result of $c_h = 0.35$ qualitatively reproducing best the observed star-like pattern at every 2$^{nd}$ BZ center. This is shown in Fig. S9(a) without form-factor and Fig. 9(b) with the Ru$^{3+}$ form factor. Comparing the result to the data in Fig. 4(a), we conclude that the RPA calculations start to correctly capture some aspects of the Q-dependence of the data beyond the pure Kitaev model, in particular the more pronounced star-like structure in the BZ. We note that a full RPA calculation would start from a model Hamiltonian and properly treat the small terms of the pure Kitaev calculation in a perturbative fashion, however the present mean-field approximation gives a general idea of the results that one should expect.

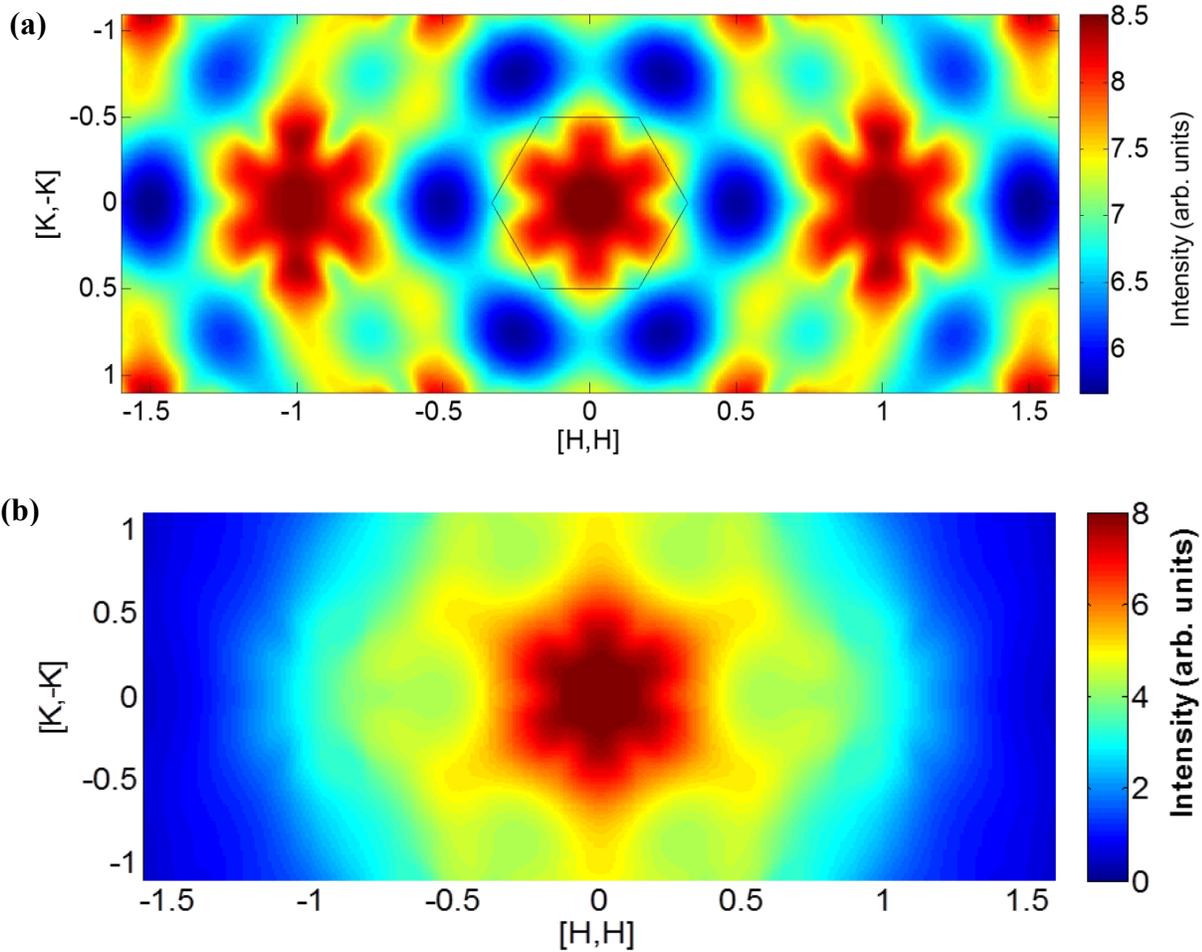

**Figure S9:** The results of the RPA calculation with $c_h = 0.35$ as described in main text **(A)** without and **(B)** with the $Ru^{3+}$ form factor for $C_l = 1$.

**F. High frequency vs. low frequency dynamics**

Evaluations of various spin-wave theory (SWT) descriptions for $\alpha$-RuCl$_3$ have been considered previously (24). In principle SWT can give a good description of the low energy excitations in a system with long range order even when the high energy spectrum is dominated by quantum fluctuations. A classic example is provided by coupled $S=1/2$ antiferromagnetic Heisenberg chains, with experiments exemplified by KCuF$_3$ (see, for e.g., Ref (44)). The situation in $\alpha$-RuCl$_3$ is believed to be similar, in the sense that spin wave excitations are present at low energies below $T_N$ while the fractionalized excitation continuum persists at high energies for T above and below $T_N$.

Most estimates of the low-energy effective Hamiltonian for α-RuCl$_3$ using DFT find the Kitaev interaction to be ferromagnetic, with a substantial contribution from off diagonal exchange and Heisenberg exchange to at least third neighbor spins (29, 30, 42). The starting point for the DFT is a perfectly ordered static structure, and the derived low-energy effective Hamiltonian depends sensitively on the precise values of bond angles and distances in the structure. In the actual material there can be disorder due to defects such as stacking faults; moreover there are substantial low frequency vibrational excitations including some that are quasi-2D. These deviations, from a structure that is both static and regular, raise the possibility of a modified effective Hamiltonian describing the actual physics seen by the neutrons, in particular at higher frequencies. The nature and size of such effects, including perhaps even on the sign of the Kitaev interaction, will need to be evaluated in the future (43).

**References:**

[1] L. Balents, *Nature* **464**, 199-208 (2010).

[2] P.W. Anderson, *Mat. Res. Bull.* **8**, 153 (1973).

[3] M. Barakeshli, E. Berg, S. Kivelson, *Science* **346**, 722-725 (2014).

[4] R.B. Laughlin, *Phys. Rev. Lett.* **50**, 1395 (1983).

[5] C. Castelnovo, R. Moessner, S.L. Sondhi, *Nature* **451**, 42 (2008).

[6] Y. Jompol *et al.*, *Science* **325**, 597 (2009).

[7] A. Kitaev, *Annals of Phys.* **321**, 2-111 (2006).

[8] J.G. Rau, E. K-H. Lee, H-Y Kee, *Annu. Rev. Condens. Matter Phys.* **7**, 195–221 (2015).

[9] J.G. Rau, E. K-H. Lee, H-Y. Kee, *Phys. Rev. Lett.* **112**, 077204 (2014).

[10] G. Baskaran, S. Mandal, R. Shankar, *Phys. Rev. Lett.* **98**, 247201 (2007).

[11] J. Knolle, D.L. Kovrizhin, J.T. Chalker, R. Moessner, *Phys. Rev. Lett.* **112**, 207203 (2014).

[12] J. Knolle, D.L. Kovrizhin, J.T. Chalker, R. Moessner, *Phys. Rev. B* **92**, 115127 (2015).

[13] A.Yu. Kitaev, *Annals of Phys.* **303**, 1 (2003).

[14] C. Nayak *et al.*, *Rev. Mod. Phys.* **80**, 1083-1159 (2008).

[15] G. Jackeli, G. Khaliullin, *Phys. Rev. Lett.* **102**, 017205 (2009).

[16] S.K. Choi *et al.*, *Phys. Rev. Lett.* **108**, 127204 (2012).

[17] S.H. Chun *et al.*, *Nature Phys.* **11**, 462 (2015).

[18] Modic, K. A. *et al.*, *Nature Comm.* **5**, 4203 (2014).

[19] A. Biffin, *et al.*, *Phys. Rev. Lett.* **113**, 197201 (2014).

[20] K.W. Plumb *et al.*, *Phys. Rev. B* **90**, 041112(R) (2014).

[21] R.D. Johnson, *et al.*, *Phys. Rev. B* **92**, 235119 (2015).

[22] J. Nasu *et al.*, *Nature Phys.* (2016), http://dx.doi:10/1038/nphys3809.

[23] L.J. Sandilands *et al., Phys. Rev. Lett.* **114**, 147201 (2015).

[24] A. Banerjee, *et al.*, *Nature Mat.* **15**, 733 (2016).

[25] H.B. Cao *et al.*, *Phys. Rev. B* **93**, 134423 (2016).

[26] J. Nasu, M. Udagawa, Y. Motome, *Phys. Rev. Lett.* **113**, 197205 (2014).

[27] J. Nasu, M. Udagawa, Y. Motome, *Phys. Rev. B* **92**, 115122 (2015).

[28] D.J. Scalapino, Y. Imry, P. Pincus, *Phys. Rev. B* **11**, 2042 (1975).

[29] H-S. Kim, H-Y. Kee, *Phys. Rev. B* **93**, 155143 (2016).

[30] S.M. Winter *et al.*, *Phys. Rev. B* **93**, 214431 (2016).

[31] J. A. Sears *et al.*, Phys. Rev. B **91**, 144420 (2015).

[32] M. Majumder *et al.*, Phys. Rev. B **91**, 180401(R) (2015).

[33] K. Brodersen *et al.*, J. Less-Common Mat. **15**, 347 (1968).

[34] M. B. Stone *et al.*, *Rev. Sci. Instrum.* **85**, 045113 (2014).

[35] G.E. Granroth *et al.*, *J. Phys.: Conference Series* **251**, 012058 (2010).

[36] O. Arnolda, *et al.*, *Nuclear Instruments and Methods in Physics Research* **764,** 156-166 (2014).

[37] R.T. Azuah, *et al., J. Res. Natl. Inst. Stan. Technol.* **114,** 341 (2009).

[38] R.A. Ewings *et al.*, http://arxiv.org/abs/1604.05895 (2016).

[39] S. Toth, B. Lake, *J. Phys. Condens. Matter* **27**, 166002 (2014).

[40] G.L. Squires, *'Introduction to the Theory of Thermal Neutron Scattering'*, 3[rd] edition, University of Cambridge Press (2012).

[41] X.-Y. Song, Y.-Z. You, and L. Balents, *Phys. Rev. Lett.* **117**, 037209 (2016).

[42] R. Yadav *et al.,*(2016) http://arxiv.org/1604.04755v1

[43] J. Chaloupka, G. Khalliullin, (2016), http://arxiv.org/1607.05676v1

[44] B. Lake, A. Tennant, C.D. Frost, S.E. Nagler, *Nature Mat.* **4**, 329 - 334 (2005).

―――――――